\documentclass[prd,aps,twocolumn,nofootinbib]{revtex4}
\usepackage{url}
\usepackage{natbib}
\usepackage{color}
\usepackage{amsmath}
\usepackage{graphicx}
\usepackage{comment}
\begin{document}

\title{Baryon asymmetry constraints on magnetic field from the Electroweak epoch}

\author{Theo Boyer$^1$, Andrii Neronov$^{1,2}$ }

\affiliation{$^{1}$Universit\'e Paris Cit\'e, CNRS, Astroparticule et Cosmologie, 
F-75013 Paris, France}

\affiliation{$^{2}$Laboratory of Astrophysics, \'Ecole Polytechnique F\'ed\'erale de Lausanne, CH-1015 Lausanne, Switzerland}

\begin{abstract}
Decay of helical (hyper)magnetic fields that may have been present in the Universe during the Electroweak epoch can contribute to generation of the baryon asymmetry of the Universe. We revise constraints on the strength and correlation length of such fields from the requirement that their decay does not lead to over-production of the baryon asymmetry. We show that the helical fields with strength down to $\sim 10^{-5}$ of the maximal possible strength during the Electroweak epoch should have had their correlation at least $\sim 10^{-6}$ of the Hubble radius during this epoch. For weaker fields this lower bound on the correlation length relaxes proportionally to the square of magnetic field strength. A field with parameters saturating the bound may actually  be responsible for  the baryon asymmetry observed today. We show that relic of such a field, surviving in the present day Universe in the form of intergalactic magnetic field detectable with Cherenkov Telescope Array Observatory, may have the strength up to 10-100~pG and can have parameters needed to affect the cosmological recombination and relax the Hubble tension. We also show that there is no constraint on the parameters of helical or non-helical magnetic fields stemming from the  requirement that the baryon isocurvature perturbations produced by such fields during the Electroweak epoch are within the observational limits.
\end{abstract}
\maketitle

Gamma-ray observations have established the existence of Inter-Galactic Magnetic Fields (IGMF) in the voids of the Large Scale Structure of the Universe \cite{2010Sci...328...73N,MAGIC:2022piy,HESS:2023zwb}, with a conservative lower bound on the field strength at the level \cite{MAGIC:2022piy} $B \gtrsim 10^{-17} \mathrm{G}$, or stronger bounds up to $B\gtrsim (2..4)\times 10^{-14}$~G that depend on assumptions about intrinsic properties of $\gamma$-ray sources used to probe the intergalactic medium along lines of sight to these sources \cite{HESS:2023zwb}.  The existing state-of-art of modeling of astrophysical processes producing such IGMF in the low-redshift Universe suggests that the fields spread  by galactic outflows are perhaps not strong enough to fill the voids \cite{Bondarenko:2021fnn,Blunier:2024aqx}, so that the void IGMF is possibly of  cosmological origin \cite{Durrer:2013pga,Vachaspati:2020blt}.

One possible mechanism of generation of cosmological magnetic fields is through quantum anomalies \cite{Joyce:1997uy} that link changes of baryon and lepton numbers to the changes in helicity of hyper-magnetic fields before the Electroweak phase transition. This is an interesting possibility, because it links the cosmological magnetic field (whose relic may be the  void IGMF today) to the baryon asymmetry of the Universe, thus providing an observational constraint on on the mechanism of generation of this asymmetry \cite{Triangle_anomaly,Giovannini:1997gp,Fujita:2016igl,Kamada:2016_1,Kamada:2016_2}. 

The link between the present-day magnetic field and the baryon asymmetry has been explored quantitatively in a series of papers \cite{Fujita:2016igl,Kamada:2016_1,Kamada:2016_2,Kamada:2020bmb}. These publications have solved numerically kinetic equations following the evolution of hyper-magnetic field together with asymmetries in concentrations of the Standard Model particles  and antiparticles through the Electroweak transition, taking into account interactions between particles and hyper-magnetic fields before the Electroweak transition and conversion of the hyper-magnetic field into magnetic field at the Electroweak transition.  Ref. \cite{Fujita:2016igl} has found that it is possible that the baryon asymmetry of the Universe has been generated by the decaying hypermagnetic field leaving the relic field in the present-day Universe with parameters consistent with the lower bound on the IGMF from $\gamma$-ray observaitons.  To the contrary, Ref. \cite{Kamada:2016_1} has found that the observed level of the baryon asymmetry of the Universe $\eta_b\sim 10^{-10}$ cannot be produced by the decaying hypermagnetic helicity, no matter what are the parameters of the magnetic field. Finally, Ref. \cite{Kamada:2016_2} has reached an opposite conclusion: that magnetic fields with present day relic strength comparible with the lower bound from $\gamma$-ray observations would typically over-produce the baryon asymmetry and hence the observed IGMF cannot be helical magnetic field from the Electroweak epoch. 

The baryon number $\eta_b$ generated by the decaying hyper-magnetic helicity is variable in space \cite{Giovannini:1997gp}. The level of spatial fluctuations of the baryon number density $S_b(\vec x)=\delta\eta_b(\vec x)/\langle\eta_b\rangle$
is constrained by the non-observation of the effect of such fluctuations on the dynamics of the Big Bang Nucleosynthesis  \cite{Inomata:2018htm}. Ref. \cite{Kamada:2020bmb} has used this bound to derive extremely tight constraints on both helical and non-helical IGMF that might originate from the Electroweak epoch, reaching as low as $B\lesssim 10^{-23}$~G for the magnetic field nearly homogeneous over the Hubble volume. 

In what follows we revise the results of Refs. \cite{Kamada:2016_1,Kamada:2016_2,Kamada:2020bmb}. First, we relax the assumption that the hyper-magnetic field present in the Universe at the Electroweak epoch is not generated during this epoch but it is a pre-existing field originating from earlier times in the history of the Universe. We include a possibility that the field present during the Electroweak transition is forced by some (unspecified) mechanism during this epoch. Next, we notice that the estimate of Ref.  \cite{Kamada:2016_2,Kamada:2020bmb} based on a specific model of additional baryon number production through the transformation of hyper-magnetic field into magnetic field suffers from an ambiguity. Finally, we consider recent developments in the modeling of evolution of turbulent decay of magnetic fields \cite{Hosking:2022umv,Brandenburg:2024tyi} that show that the decay  proceeds slower than suggested by a conventional back-of-the-envelope estimates, so that the largest turbulently processed eddies have smaller size, compared to the model of Ref. \cite{Banerjee:2004df} that has been used in the analysis of Refs. \cite{Kamada:2016_1,Kamada:2016_2,Kamada:2020bmb}. We show that this leads to a larger efficiency of baryon number generation by the decaying helicity of hyper-magnetic field that might have been produced during epochs preceding the Electroweak transition. 
 
\section{Summary of the methodology}

Within the framework of Refs. \cite{Kamada:2016_1,Kamada:2016_2,Kamada:2020bmb},  the dynamics of the baryon number generation is described by a system of coupled kinetic equations  \cite{Kamada:2016_1}
\begin{equation}
\frac{d\vec{\eta}}{d x} =\mathbf{M} \vec{\eta}+\overrightarrow{\mathcal{S}}
\label{eq:kinetic}
\end{equation}
describing the evolution of asymmetries $\eta=(n-\overline n)/s$ (differences in the number densities of particles $n$ and antiparticles $\overline n$ divided by the entropy density $s$) of different left- and right-handed species (leptons and quarks, $W^\pm$ bosons, the Higgs field).  The derivatives in the left hand side of Eqs. (\ref{eq:kinetic})  is taken with respect to the temperature coordinate $x=M_0/T$ where  $M_0\simeq 7\times 10^{17}$~GeV. The vector $\vec S$ contains all the source terms that depend on the (hyper)magnetic field, while the matrix $\mathbf{M}$ contains all the transport coefficients corresponding to all the interactions between particles. We refer to \cite{Kamada:2016_1} for the detailed formulation of the system of equations (\ref{eq:kinetic}).

It has been shown in Ref. \cite{Kamada:2016_1} that solutions of equations (\ref{eq:kinetic}) closely follow a quasi-equilibrium state in which the source terms balance the "wash-out" terms, so that the two terms in the right hand side of Eqs. (\ref{eq:kinetic}) nearly cancel each other. Different quasi-equilibrium solutions can be found for different temperature intervals: before the Electroweak transition, $T\gtrsim 160$~GeV and  in the temperature range $T\lesssim 160$~GeV, down to the temperature $T_*\simeq 130$~GeV of freeze-out of sphaleron interactions. 

The peculiarity of the quasi-equilibrium solution in the $130$~GeV$<T<160$~GeV range is that even though the magnetic field does not seed the baryon asymmetry anymore, the wash-out of this asymmetry is slowed down by the weakness of the Yukawa couplings that hamper transfer of the asymmetry between the right and left-handed particles. This leads to a non-zero value of $\eta_b$ at $T\simeq T_*$. 

It was noticed in Ref. \cite{Kamada:2016_2} that the conversion of hyper-magnetic field into the magnetic field after the electroweak phase transition may happen not instantaneously at $T\simeq 160$~GeV, but rather gradually over a wider temperature interval. In this case, the hypermagnetic field may still keep sourcing some baryon asymmetry during the time between the moment when the Higgs field gets its vacuum expectation value  and the moment of the sphaleron freeze-out.  In this case,  this additional source modifies the quasi-equilibrium value of $\eta_b$, boosting it by an uncertain, potentially large, factor. 

To model this boost,  Refs. \cite{Kamada:2016_2,Kamada:2020bmb} have introduced a phenomenological model with an effective $U(1)$ field with potential  ${\cal A}$ related to the Standard Model  $Y$ and $W$ field potentials as 
$Y_\mu(t,\vec x)=\cos\theta_w(t){\cal A}_\mu(\vec x)$, $W_\mu^3(t,\vec x)=\sin\theta_w(t){\cal A}_\mu(\vec x)$,
where $\theta_w (t)$ is a function that interpolates between $0$ at the moment of the Electroweak transition and the Weinberg angle at a later time, possibly reaching its final value only when the temperature drops down to $T=T_*$.  The shape of the function $\theta_w(t)$ has been estimated from lattice simulations \cite{DOnofrio:2015gop}, but the shape of the effective time-independent part ${\cal A}_\mu(\vec x)$ has not been derived from dynamical equations and it is not clear if the proposed separation of time and space dependence of the fields $Y_{\mu}, W^3_\mu$ is possible. 
We notice that the ${\cal A}_\mu(\vec x)$ is not a gauge field in the sense that  the theory  is not gauge invariant with respect to $U(1)$ gauge transformations of this field, ${\cal A}_\mu\rightarrow {\cal A}_\mu+\partial_\mu f$. The observable of interest, the spatially variable  baryon asymmetry,  $\eta_b (\vec x)\propto {\cal A}(\vec x){\cal B}(\vec x)$ \cite{Kamada:2020bmb} (${\cal B}$ is the strength of effective magnetic field associated to ${\cal A}$), changes under the gauge transformations of ${\cal A}$. The choice of the funciton $f$ has not been discussed in Ref. \cite{Kamada:2016_2}, which assumed that the result would not depend on the "gauge fixing". However, choosing $f(\vec x)$ variable with amplitude $f_0$ on a distance scale $\lambda$, $f(x)\sim f_0(x/\lambda)$, one can change $\eta_b$ by an amount $\delta \eta_b\propto f_0 {\cal B}/\lambda$ which can be arbitrarily large, depending on the value of $f_0$, so that it not possible to make estimates of $\eta_b$ and its fluctuations on any distance scale $\lambda$ in the absence of a model prescription for fixing the shape of $f$.

Given this uncertainty, we adopt the approach of Ref. \cite{Kamada:2016_1} in which the possibility of additional injection of the baryon asymmetry by this conversion was not considered. This approach might somewhat underestimate the resulting baryon asymmetry. This makes the constraints on magnetic field parameters presented below  "conservative", in the sense that a proper modeling of conversion of hypermagnetic to magnetic field may strengthen, but not weaken these constraints. 

\section{Results}

\subsection{Bound on the strength and correlation length of helical magnetic field at the Electroweak epoch}

Within the setup of Ref. \cite{Kamada:2016_1},  the final value of the baryon asymmetry of the Universe can be derived from the approximate solution of the kinetic equations (\ref{eq:kinetic}) at the temperature $T\simeq T_*$. This solution reads
\begin{equation}
\label{eq:equilibrium}
    \eta_b\simeq \frac{11}{37}\frac{S_{em}}{\gamma_{h\leftrightarrow ee}+\gamma_{flip}+\gamma_{em}^{CME}}
\end{equation}
where the source term is\footnote{We use the Natural system of Units $c=\hbar=k_B=1$.}
\begin{equation}
    S_{em}\simeq 10^{-6}\frac{\langle{\vec B\cdot\vec \nabla\times \vec B}\rangle}{ T_*^5}
\end{equation}
with  $g_*$ being the number of relativistic degrees of freedom and the averaging is over the volume. The transport coefficient $\gamma_{em}^{CME}$ depends  on the magnetic field
\begin{equation}
\gamma_{em}^{\mathrm{CME}} = 0.1\frac{\alpha_{em}^2}{\pi^2} \frac{\langle\vec{B}^2\rangle}{T_*^4}\simeq 5\times 10^{-7} \frac{\langle\vec{B}^2\rangle}{T_*^4}
\end{equation}
($\alpha_{em}$ is the fine structure constant), while the  other transport coefficients are magnetic field independent:
\begin{eqnarray}
    &&\gamma_{h\leftrightarrow ee}\simeq \frac{6\ln(2)|y_{e}^{11}|^2}{8\pi^3}\frac{m_h(T)^2}{T^2}\simeq 8\times 10^{-14} \nonumber \\
    &&\gamma_{flip}\simeq 10^{-2}\frac{|y_{e}^{11}|^2}{\pi^2}\frac{v(T)^2}{T^2}\simeq 2\times 10^{-14}
\end{eqnarray}
with the  the right electron Yukawa coupling $|y_e^{11}|\simeq 3\times 10^{-6}$, Higgs mass and vacuum expectation value $m_h(T)\simeq 90$~GeV and $v(T)\simeq 170$~GeV at the temperature $T\simeq T_*$.

\begin{figure}
\includegraphics[width=\columnwidth]{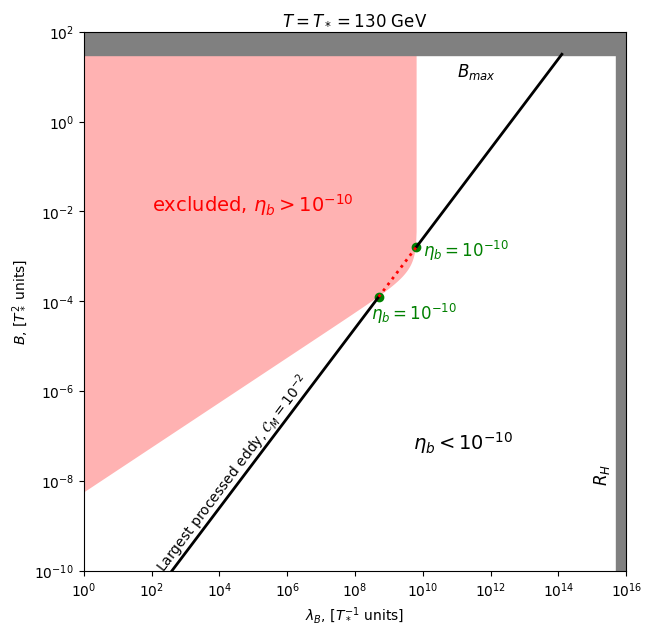}
\caption{Constraint on the strength and correlation length of helical magnetic field in the Universe at the temperature $T_*=130$~GeV, imposed by the requirement that the baryon asymmetry $\eta_b$ does not exceed the observed level of $10^{-10}$ (red shading). The upper limit on the magnetic field strength (gray shading) corresponds to the field with energy density in equipartition with the overall energy density of the Universe at this temperature. Upper limit on the correlation length (also gray shading) is the Hubble radius at $T=T_*$. Black solid and red dashed line corresponds to the size of the largest turbulently processed eddies in the case ${\cal C}_M=10^{-2}$. Red part of the line is excluded by the baryon asymmetry constraint. Green points correspond to the $B,\lambda_B$ values that provide the observed $\eta_b$.    }
\label{fig:constraint_initial}
\end{figure}

One can see that as long as 
\begin{equation}
\langle\vec{B}^2\rangle^{1/2}\gg 5\times 10^{-4}T_*^2,
\end{equation}
 the denominator of Eq. (\ref{eq:equilibrium}) can be simplified to yield 
\begin{equation}
\label{eq:equilibrium1}
    \eta_b\simeq 0.6\frac{\langle \vec B\cdot \vec \nabla \times \vec B\rangle}{T_*\langle \vec B^2\rangle}\sim 0.6(\lambda_B T_*)^{-1}
\end{equation}
where the order-of-magnitude estimate is for a maximally helical magnetic field for which  $\langle \vec{B}\rangle^2 \sim B^{2}$ and $\langle \vec B\cdot\vec \nabla\times \vec B\rangle  \approx B^{2}/\lambda_B$ with $B$ and $\lambda_B$ being the strength and correlation length of the field and we have used an estimate $g_*\simeq 10^2$ during the Electroweak epoch.

The equation (\ref{eq:equilibrium1}) suggests that there exists a lower bound on the correlation length of strong  helical magnetic field, stemming from the requirement that the baryon asymmetry should not exceed the observed level $\eta_b\simeq 10^{-10}$:
\begin{equation}
\lambda_B\gtrsim 10^{10}T_*^{-1}.
\end{equation}

In the opposite case, $B\ll 5\times 10^{-4}T_*^2$, Eq. (\ref{eq:equilibrium}) takes the form 
\begin{equation}
\eta_B\simeq 3\times 10^6 \frac{\langle \vec B\cdot \vec \nabla\times \vec B\rangle}{T_*^5}
\end{equation}
and the requirement $\eta_b\le 10^{-10}$ imposes a bound on a combination of the strength and correlation length of the field:
\begin{equation}
\lambda_B\gtrsim 10^{10}T_*^{-1}\left[\frac{B}{5\times 10^{-4}T_*^2}\right]^2
\end{equation}
Overall, the requirement that the decay of hypermagnetic field does not lead to over-produciton of the baryon asymmetry imposes a constraint on the strength and correlation length of helical magnetic field at the moment of the sphaleron freeze-out:
\begin{equation}
\label{eq:limits}
\lambda_B\gtrsim\left\{
\begin{array}{ll}
10^{10}T_*^{-1} &, B\gg 5\times 10^{-4}T_*^2\\
10^{10}T_*^{-1}\left(\frac{\displaystyle B}{\displaystyle5\times 10^{-4}T_*^2}\right)^2&, B\ll 5\times 10^{-4}T_*^2
\end{array}
\right.
\end{equation}
This excluded range of parameter space of the helical magnetic field at the moment  $T\sim T_*$ is shown by a semi-transparent red shading in Fig. \ref{fig:constraint_initial}. The  boundary of the excluded region corresponds to the combinations of $B$, $\lambda_B$ for which the baryon asymmetry resulting from the decay of helical hypermagnetic field is at the level observed today\footnote{The additional injection of the baryon asymmetry from the hypermagnetic-to-magnetic field conversion, discussed in the previous section, would shift the boundary to the right in Fig. \ref{fig:constraint_initial}.}. 

The upper limit on the field strength (gray shading) in Fig. \ref{fig:constraint_initial} corresponds to the strongest possible magnetic field with the energy density comparable to the overall energy density of the Universe: 
\begin{equation}
\label{eq:bmax}
    \frac{B_{max}^2}{8\pi}\sim \frac{\pi^2}{30}g_* T_*^4\simeq 30T_*^4
\end{equation}
The upper limit on the correlation length, also shown by the gray shading in Fig. \ref{fig:constraint_initial}, corresponds to the Hubble radius at $T=T_*$:
\begin{equation}
\label{eq:hubble}
R_H=\frac{M_0}{T_*^2}\sim 5\times 10^{15}T_*^{-1}
\end{equation}

\subsection{Bound on the field produced before the Electroweak epoch.}

It is possible that the hyper-magnetic field present during the Electroweak epoch originates from an earlier epoch. In this case, the pre-existing field should have experienced a turbulent decay during which both the strength and correlation length of the filed have evolved. The details of this evolution are not entirely clear. There exist several phenomenological models of evolution of magnetic field in the early universe under the influence of the free turbulence decay. In all these models the field strength and correlation length are given by a "largest processed eddy" relation:
\begin{equation}
\label{eq:result2}
\lambda_B={\cal C}_M v_A R_H \sim 5\times 10^{15} T_*^{-1}{\cal C}_M \left[\frac{B}{B_{max}}\right]
\end{equation}
where $v_A\simeq B/B_{max}$ is the Alfven velocity and ${\cal C}_M$ is a proportionality coefficient that differs from model to model. In the model of Ref. \cite{Banerjee:2004df}, ${\cal C}_M=1$, while numerical and analytical modeling of Refs. \cite{Hosking:2022umv,Brandenburg:2024tyi} find smaller values of $10^{-4}\lesssim {\cal C}_M\lesssim 0.1$. Given this modeling uncertainty, we leave ${\cal C}_M$ as a free parameter.  The scaling of Eq.  (\ref{eq:result2}) is illustrated by the inclined black and red  line in Fig. \ref{fig:constraint_initial}. Comparing the estimate of $B, \lambda_B$ from Eq. (\ref{eq:result2}) with the estimates of Eq. (\ref{eq:limits}), one can find that the requirement $\eta_b\le 10^{-10}$ imposes bounds on the possible field strength (or correlation length). Either the field should be stronger than 
\begin{equation}
B\gtrsim 6\times 10^{-3}T_*^2\left[\frac{{\cal C}_M}{10^{-2}}\right]^{-1}
\end{equation}
or it should be weaker than
\begin{equation}
B\le 4\times 10^{-5}T_*^2 \left[\frac{{\cal C}_M}{10^{-2}}\right], 
\end{equation}
where  have normalized the ${\cal C}_M$ to the value found in numerical modeling of Ref. \cite{Brandenburg:2024tyi}. The allowed part of the "largest processed eddy" line is shown by black color in Fig. \ref{fig:constraint_initial}. The part excluded by the requirement $\eta_b\le 10^{-10}$ is shown by the red dotted line. Green points at the end of the allowed lines show the values of $B,\lambda_B$ that provide correct $\eta_b$ in the present-day Universe. Such values exist in the models with ${\cal C}_M\lesssim 0.03$.

\begin{figure}
    \includegraphics[width=\columnwidth]{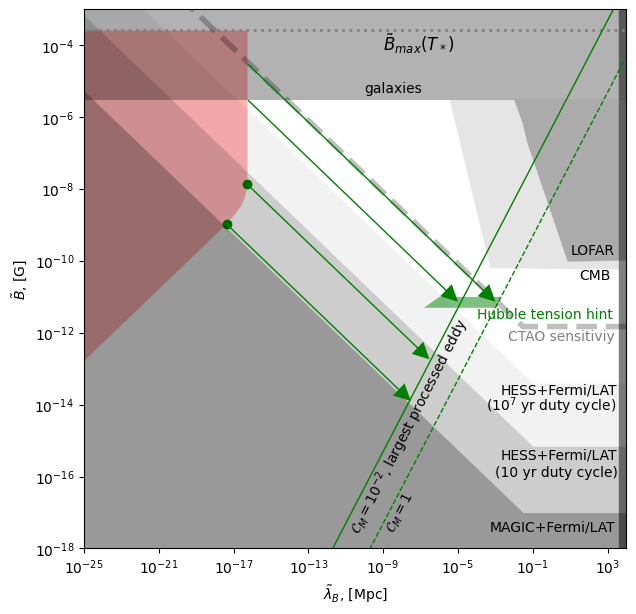}
    \caption{Parameter space of cosmological magnetic field. Red shaded region and the green circles are the same as in Fig. \ref{fig:constraint_initial}, but expressed through the co-moving magnetic field strength and correlation length. Green arrows show the evolutionary path of the field parameters up to the present day IGMF. Green solid line shows the locus of the evolution endpoints for ${\cal C}_M=10^{-2}$, green dotted line is for ${\cal C}_M=1$. Grey shading shows known observational constraints on the cosmological IGMF. The lower bounds are from MAGIC$+$Fermi/LAT $\gamma$-ray telescope  measurements \cite{MAGIC:2022piy} and a combination of HESS and Fermi/LAT  data \cite{HESS:2023zwb}. The upper bounds are from LOFAR radio telescope data \cite{Neronov:2024qtk} and from the analysis of the effect of baryon clumping on CMB \cite{Jedamzik:2018itu}. Green shaded range shows the parameters of magnetic field that relaxes the Hubble tension \cite{Jedamzik:2025cax}. The shortest possible correlation length for this region is from \cite{Hosking:2022umv}. Dashed gray line shows the sensitivity limit of CTAO from \cite{Korochkin:2020pvg}.  }
    \label{fig:constraint}
\end{figure}

\section{Present-day helical IGMF that can be responsible for the baryon asymmetry}

The helical magnetic field with parameters that provide the observed level of the baryon asymmetry of the Universe is expected to evolve through free turbulence decay up to recombination. The co-moving strength, $\tilde B=a^2 B$ ($a$ is the scale factor of the Universe), and correlation length $\tilde \lambda_B=\lambda_B/a$ of the field follow a path $\tilde B\propto \tilde \lambda_B^{-1/2}$ shown by the arrows in Fig. \ref{fig:constraint}, up to the evolution endpoints \cite{Banerjee:2004df,Hosking:2022umv,Brandenburg:2024tyi}
\begin{equation}
\label{eq:largest_processed}
\tilde \lambda_B\sim 10^{-2}\left[\frac{{\cal C}_M}{10^{-2}}\right]\left[\frac{\tilde B}{3\times 10^{-9}\mbox{ G}}\right]\mbox{ Mpc}
\end{equation}
The maximal possible initial field with the strength $\tilde B=\tilde B_{max}\simeq 2\times 10^{-4}$~G and correlation length $\tilde \lambda_B=10^{10}T_*^{-1}/a\simeq 10^{-16}$~Mpc would therefore evolve into a field with the present-day strength
\begin{equation}
    B_{max,IGMF}\sim 10^{-10}\left[\frac{{\cal C}_M}{10^{-2}}\right]^{-1/3}\mbox{ G}
\end{equation}
and correlation length 
\begin{equation}
    \lambda_{B,IGMF}\sim 5\times 10^{-4}\left[\frac{{\cal C}_M}{10^{-2}}\right]^{2/3}\mbox{ Mpc}
\end{equation}
From Fig. 1  one can see that such a field would be at the level of the maximal possible field saturating the upper bound from the observations of Cosmic Microwave Background (CMB) derived in Ref. \cite{Jedamzik:2018itu}. 

The evolutionary tracks of the field responsible for the production of   the baryon asymmetry at the Electroweak transition and originating from an earlier epoch are shown by the two lower green arrows in Fig. \ref{fig:constraint} for ${\cal C}_M=10^{-2}$. In this case the present-day relic of this field residing in the voids of the Large Scale Structure is expected to have strength $B\simeq 2\times 10^{-13}$~G and correlation length $\lambda_B\sim 3\times 10^{-7}$~Mpc. Such field parameters are below the lower bound derived from a combination of observations by HESS and Fermi $\gamma$-ray telescopes \cite{HESS:2023zwb} (light gray shading in Fig. \ref{fig:constraint}),  but above a more conservative limit derived from model-independent analysis of MAGIC and Fermi telescope data \cite{MAGIC:2022piy}.

\section{Constraint  from baryon isocurvature perturbations}

The baryon asymmetry produced by the decaying hypermagnetic helicity is in general variable in space, because $B$ is spatially variable. Let us consider the "strong field" regime 
$B\gtrsim 5\times 10^{-4}T_*^2$. 
In this case $\eta_b$ is approximated as 
\begin{equation}
\label{eq:equilibrium2}
    \eta_b\sim \frac{\vec B\cdot \vec \nabla \times \vec B}{T_* \vec B^2}
\end{equation}
We are interested in spatial fluctuations of $\eta_b$. Following \cite{Kamada:2020bmb}, we define $\delta \eta_b=\eta_b(\vec x)-\overline\eta_b$,  
\begin{equation}
S_b(\vec x)=\frac{\delta\eta_b(\vec x)}{\langle\eta_b \rangle}
\end{equation}
and the two-point correlation function of the baryon number fluctuations
\begin{equation}
{\cal G}(\vec r)=\langle S_b(\vec x)S_b(\vec x +\vec r)\rangle=\frac{\langle \eta_b(\vec x)\eta_b(\vec x+\vec r)\rangle}{\langle\eta_b^2\rangle}-1
\end{equation}
To calculate this correlation function, we need to evaluate
\begin{equation}
\langle \eta_b(\vec x)\eta_b(\vec x+\vec r)\rangle=\frac{1}{T_*^2}\left< \frac{\vec B \cdot \nabla \times \vec B}{\vec B^2}(\vec x)\frac{\vec B\cdot\nabla \times \vec B}{\vec B^2}(\vec x+\vec r)\right>
\end{equation}
To do this, we use the approach of Ref. \cite{Triangle_anomaly}:
\begin{eqnarray}
&&\left< \frac{\vec B \cdot \nabla \times \vec B}{\vec B^2}(\vec x)\frac{\vec B\cdot\nabla \times \vec B}{\vec B^2}(\vec x+\vec r)\right>\simeq\nonumber\\
&&\frac{\langle(\vec B\cdot\nabla\times \vec B)(\vec x)(\vec B\cdot\vec\nabla\times\vec B)(\vec x+\vec r)\rangle}{\langle|\vec B|^2\rangle^2}
\end{eqnarray}
so that
\begin{equation}
\langle \eta_b(\vec x)\eta_b(\vec x+\vec r)\rangle\simeq \frac{\langle(\vec B\cdot\nabla\times \vec B)(\vec x)(\vec B\cdot\vec\nabla\times\vec B)(\vec x+\vec r)\rangle}{T_*^2\langle |\vec B|^2\rangle^2}
\end{equation}
The Fourier transform of ${\cal G}(\vec r)$ can be written as \cite{Kamada:2020bmb}
\begin{eqnarray}
&&\mathcal{G}(\vec{k})  =\frac{1}{V} \frac{\left\langle \left| \eta_{B}(\vec{k})\right|^2\right\rangle}{\bar{\eta}_B^2}-(2 \pi)^3 \delta^3(\vec{k}) \nonumber\\
&& =\epsilon_{i j k} \epsilon_{l m n} \frac{\mathcal{C}^2}{V \bar{\eta}_B^2} \int \frac{d^3 p}{(2 \pi)^3} \int \frac{d^3 p^{\prime}}{(2 \pi)^3} \nonumber\\ &&p_j p_m^{\prime} J_{i k l n}\left(\vec{k}-\vec{p}, \vec{p}, \vec{k}-\vec{p}^{\prime}, \vec{p}^{\prime}\right)-(2 \pi)^3 \delta^3(\vec{k})
\end{eqnarray}
where 
\begin{eqnarray}
&&J_{i k l n}\left(\vec{k}_1, \vec{k}_2, \vec{k}_3, \vec{k}_4\right)=\left\langle B_i^*(\vec{k}_1) B_k^*(\vec{k}_2) B_l(\vec{k}_3) B_n(\vec{k}_4)\right\rangle \nonumber\\
&& =\left\langle B_i^*(\vec{k}_1) B_k^*(\vec{k}_2)\right\rangle\left\langle B_l(\vec{k}_3) B_n(\vec{k}_4)\right\rangle+\nonumber\\ 
&&\left\langle B_i^*(\vec{k}_1) B_l(\vec{k}_3)\right\rangle\left\langle B_k^*(\vec{k}_2)B_n(\vec{k}_4)\right\rangle+ \nonumber\\
&& \left\langle B_i^*(\vec{k}_1) B_n(\vec{k}_4)\right\rangle\left\langle B_k^*(\vec{k}_2) B_l(\vec{k}_3)\right\rangle
\end{eqnarray}
and
\begin{equation}
\left\langle B_{i}^*(\vec{k}) B_{j}(\vec{k}^{\prime})\right\rangle=(2 \pi)^3 \delta^3\left(\vec{k}-\vec{k}^{\prime}\right) \tilde{\mathcal{F}}_{i j}^B(\vec{k})
\end{equation}
Similar to \cite{Kamada:2020bmb}, we decompose the spectrum  $\tilde{\mathcal{F}}_{i j}^B(\vec{k})$  into symmetric and an antisymmetric parts:
\begin{equation}
\tilde{\mathcal{F}}_{i j}^B(\vec{k})=P_{i j}({\vec{k}}) \tilde{S}^B(k)+i \epsilon_{i j m} {k}_m \tilde{A}^B(k)
\end{equation} 
where $P_{ij}({\vec{k}}) = \delta_{ij}-{k}_i {k}_j$. Considering magnetic field with helicity fraction $\epsilon$, 
\begin{equation}
A^B(k)=\epsilon S^B(k)
\label{complete expression}
\end{equation} one finds
\begin{eqnarray}
&&{\cal G}(\vec k)=\frac{1}{\bar{\eta}_B^2 T_*^2\langle |\vec B^2|\rangle^2} \\
&&=\int \frac{d^3 p}{(2 \pi)^3}S^B(|\vec{k}-\vec{p}|) S^B(p)\left[p^2 +\epsilon^2|\vec{k}-\vec{p}| p\right] \nonumber\\
&& \times\left[1-\frac{2(\vec{k}-\vec{p}) \cdot \vec{p}}{p^2}+\frac{((\vec{k}-\vec{p}) \cdot \vec{p})^2}{|\vec{k}-\vec{p}|^2 p^2}\right]\nonumber
\end{eqnarray}
The spectrum of baryon density fluctuations gets suppressed on the distance scales shorted than the neutrino diffusion length $\lambda_n$. This suppression has been modeled in Ref. \cite{Kamada:2020bmb} by a Gaussian-type $\exp(-k^2/2D)$ suppression factor in the calculation of the overall level of the baryon density fluctuations, with $D\sim 1/\lambda_n^2$.  Overall, repeating the derivation of Ref. \cite{Kamada:2020bmb} replacing ${\cal A}$ by $B$ where necessary, we find the baryon number density fluctuations 
\begin{widetext}
\begin{eqnarray}
&& \langle{S_{b, \mathrm{BBN}}^2}\rangle= \frac{1}{4 \pi^4 \bar{\eta}_B^2T_*^2\langle |\vec B|^2\rangle^2}\int d k_1 d k_2 k_1^2 k_2^2 S^B(k_1) S^B(k_2)\nonumber\\ 
&&\left\{\frac{(k_1 + k_2)^2}{2}(1 + \epsilon^2) \frac{D}{k_1 k_2}\left(1 - \frac{D}{k_1 k_2}\right)+\left[\frac{k_1^2+k_2^2}{2} +\epsilon^2 k_1 k_2 \right]\left(\frac{D}{k_1 k_2}\right)^3\right\} \exp \left[-\frac{\left(k_1 - k_2\right)^2}{2 D}\right] \nonumber\\
&& -\left\{\frac{(k_1 - k_2)^2}{2} (1- \epsilon^2) \frac{D}{k_1 k_2}\left(1 + \frac{D}{k_1 k_2}\right)+\left[\frac{k_1^2+k_2^2}{2} +\epsilon^2k_1 k_2 \right]\left(\frac{D}{k_1 k_2}\right)^3\right\} \exp \left[-\frac{\left(k_1 + k_2\right)^2}{2 D}\right] .
\label{complet_expression_sbbn}
\end{eqnarray}
\end{widetext}
In the specific case of delta-function power spectrum with power concentrated at a specific wavenumber $k_B$, 
\begin{equation}
S^B(k)=\pi^2 \frac{B_{0}^2}{k_B^2} \delta\left(k-k_B\right)
\end{equation}
the expression for $\langle{S_{B, \mathrm{BBN}}^2}\rangle$ reduces to
\begin{eqnarray}
&& \langle{S_{B, \mathrm{BBN}}^2}\rangle= \frac{(1+\epsilon^2)}{4 \bar{\eta}_b^2T_*^2}\\ 
&&\left[D\left(1 - \frac{D}{k_B^2}\right) +k_B^2\left(\frac{D}{k_B^2}\right)^3 \left(1-\exp \left[-\frac{2k_B^2}{D}\right] .\right)\right] \nonumber
\end{eqnarray}
The  neutron diffusion distance  scale is $\lambda_n\simeq 3\times 10^5$~cm at the moment of BBN. When  blueshifted back to the Electroweak epoch  it is $\lambda_n\simeq 0.3$~cm, just somewhat smaller than  the Hubble scale. Thus, we expect $k_B^2=1/\lambda_B^2\ge D=1/\lambda_n^2$. In the  case $k_B^2\gg D$ the result simplifies to 
\begin{equation}
\langle{S_{B, \mathrm{BBN}}^2}\rangle\simeq  \frac{(1+\epsilon^2)}{4 \bar{\eta}_b^2T_*^2\lambda_n^2}\sim 10^{-11}(1+\epsilon^2)
\label{eq:small_corr_length}
\end{equation}
where we have used the observed value $\eta_b\sim 10^{-10}$ for the numerical estimate.
This is much lower than the observational constraint \cite{Inomata:2018htm}
\begin{equation}
\langle S_{b, \mathrm{BBN}}^2\rangle<0.016 \quad(2 \sigma \mbox{ level})
\end{equation}
so that effectively the requirement that the baryon isocurvature perturbations produced by the conversion of hypermagnetic field (helical or non-helical) do not exceed the observational limit does not constrain the magnetic field. In the opposite limit, $k_B^2\ll D$, the estimate for the  $\langle S_{b, \mathrm{BBN}}^2\rangle$ increases by a factor $D/k_B^2$, but the maximal possible correlation length of magnetic field can at most be comparable to $R_H$, so that this factor cannot boost the level of baryon number fluctuations by more than a factor of $\sim 10^2$ and the resulting value of $\langle S_{b, \mathrm{BBN}}^2\rangle$ is still much below the observational bound.

\section{Discussion and conclusion}

In this paper, we have revised the results of Refs. \cite{Kamada:2016_1,Kamada:2016_2,Kamada:2020bmb}, reassessing the details of evolution of the (hyper)magnetic field through the Electroweak epoch. We have shown that 
\begin{itemize}
\item  if helical fields have been present during the Electroweak phase transition, their anomalous coupling to the baryons may lead to the production of thebaryon asymmetry at the level compatible with observations in the present day Universe and that
\item  the observational constraint on the baryon isocurvature perturbations does not constrain the strength and/or correlation length of the relic helical or non-helical intergalactic magnetic fields surviving from the Electroweak epoch. 
\end{itemize}
The requirement that the baryon asymmetry $\eta_b$ resulting from the decay of the hypermagnetic helicity does not exceed the present-day level imposes a constraint on the strength and correlation length of helical magnetic field at the moment when the temperature of the Universe was $T=T_*\simeq 130$~GeV.  Helical fields with parameters in the red-shaded region of Fig. \ref{fig:constraint_initial} are ruled out. 

Given that the calculation presented above omits the possibility of additional generation of the baryon number in the case when conversion of hyper-magnetic field into magnetic field does not complete fast enough at the temperature close to the temperature of the Electroweak phase transition ($T\simeq 160$~GeV) but instead continues in the temperature interval down to $T\simeq T_*$, the constraint shown in Fig. \ref{fig:constraint_initial} is "conservative". Account of the additional injection of $\eta_b$ by the hypermagnetic-to-magnetic field conversion would only strengthen it. In any case, the boundary of the excluded region  is the locus of magnetic field parameters that provide $\eta_b$ at the level of the observed baryon asymmetry. Thus, the decay of hypermagnetic helicity can, in principle, be responsible for the generation of the observed baryon asymmetry of the Universe. 

The hyper-magnetic field seeding the baryon number may be either generated right at the Electroweak epoch or it may be originating from previous epochs in the history of the Universe. In this latter case, the field strength and correlation length are expected to be related as in Eq. (\ref{eq:result2}), with a coefficient ${\cal C}_M$ that depends on the model adopted for description of magnetized turbulence in the time range between the moment of field generation and  the Electroweak epoch. We have shown that models that predict ${\cal C}_M\lesssim 0.03$ \cite{Hosking:2022umv,Brandenburg:2024tyi} provide predictions for the evolved hypermagnetic field parameters that may be consistent with the possibility that such fields are responsible for the generation of the baryon asymmetry of the Universe (green circles for in Fig. \ref{fig:constraint_initial} show examples ofr the ${\cal C}_M= 10^{-2}$ case). 

The helical magnetic field that may be responsible for the generation of the baryon asymmetry may survive till present day in the form of the IGMF occupying the voids of the Large Scale Structure. Our analysis shows that the present-day strength of this field may reach $\sim 10^{-10}$~G, if the initial field strength at the Electroweak epoch was close to its maximal possible value. This is an interesting possibility in the view of the study of Ref. \cite{Jedamzik:2025cax} that finds that a common fit of CMB and Large Scale Structure data favors the presence of magnetic field that would leave 5-10 pG relic field in the present day Universe. This range is shown by the green shading in Fig. \ref{fig:constraint}. It is stretched along the $x$ axis within the current uncertainty of the modeling of the locus of the magnetic field evolution endpoints, with the shortest correlation length estimate from Ref. \cite{Hosking:2022umv} and largest correlation length estimate from Ref. \cite{Banerjee:2004df}. From Fig. \ref{fig:constraint} one can see that it is possible to find evolutionary tracks of helical magnetic field that connect the field that may be responsible for the generation of the baryon asymmetry to the field that would influence the recombination and relax the Hubble tension. The correlation length of such field at the moment of generation at the Electroweak epoch should have been $\lambda_B\sim 10^{10}T^{-1}$ (or some $\sim 10^{-6}$ of the Hubble radius at the Electroweak epoch), while its strength should have been $B\sim T_*^2$ (a factor of $10-100$ lower than the maximal possible field strength at the Electroweak epoch). 

The present-day relic parameters of such field are within the sensitivity reach of the new Cherenkov Telescope Array Observatory (CTAO), shown by the dashed line in Fig. \ref{fig:constraint} \cite{Korochkin:2020pvg}. Weaker magnetic fields are already probed by the current generation $\gamma$-ray telescopes (lower gray shaded regions in Fig. \ref{fig:constraint}). The most conservative lower bound on the IGMF is imposed by Fermi Large Area Telescope (LAT) non-observation of an "echo" of variable TeV band $\gamma$-ray emission from a blazar 1ES 0229+200 by MAGIC telescope \cite{MAGIC:2022piy}. 

Stronger lower bounds have been derived from the Fermi/LAT search of extended emission around blazars detected in the TeV band by HESS telescope \cite{HESS:2023zwb}. These limits depend on the assumption about stability of the blazar activity over extended time periods. Historical $\gamma$-ray observations in the TeV band now span two decades time scale and many blazars are observed to have persistent (although variable level) activity on this time scale. The bound derived under assumption of 10~yr scale duty cycle of blazar activity is more than an order-of-magnitude stronger compared to MAGIC+Fermi/LAT bound. 
This lower bound is in tension with the possibility that the present-day IGMF is the result of evolution of the hypermagnetic field that has been produced before the electroweak phase transition, has evolved into the field with parameters shown by the green circles in Figs. \ref{fig:constraint_initial} and \ref{fig:constraint} and at the time of the phase transition has generated the baryon asymmetry. It is possible to scrutinize the assumption about stability of TeV $\gamma$-ray flux from the blazar sample studied in Ref. \cite{HESS:2023zwb} via long-term monitoring of the sources of interest with CTAO and also with LHAASO $\gamma$-ray observatory \cite{LHAASO:2023rpg}. 

\section*{Acknowledgments}

We are grateful to K.Kamada and M.Shaposhnikov for clarifying discussions of the subject. The work has been supported in part by the French National Research Agency (ANR) grant ANR-24-CE31-4686.

\bibliography{refs.bib}

\end{document}